# Raman spectra of L-leucine crystals

P.F. Façanha Filho, P.T.C. Freire*, K.C.V. Lima, J. Mendes Filho, F.E.A. Melo

Departamento de Física, Universidade Federal do Ceará
Campus do Pici, C.P. 6030 Fortaleza-CE 60455-760 Brazil

P.S. Pizani

Departamento de Física, Universidade Federal de São Carlos
São Carlos – SP 13565-905 Brazil

## Abstract

*Single crystal samples of L-leucine, $C_6H_{13}NO_2$, a fundamental aliphatic amino acid of the human body, have been studied by Raman spectroscopy at temperatures from 300 to 430 K over the spectral range from 50 to 3100 cm$^{-1}$. A tentative assignment of all bands is given. For high temperatures, several modifications on the Raman spectra were observed at about 353 K, giving evidence that the L-leucine crystal undergoes a structural phase transition.*

Email: tarso@fisica.ufc.br Phone: 55.85.33669535, Fax: 55.85.33669450

## Introduction:

The vibrational and structural characterization of amino acid crystals under extreme conditions has gained much attention in the last years. This is because of the possibility of using them in technological devices, mainly in those cases where amino acid crystallizes together with other inorganic molecules, as occurs for L-arginine phosphate, which presents a high non-linear coefficient and can be used as a non-linear material involving several applications [1, 2]. Additionally, there are other aspects related to physical behavior that can be observed easily by vibrational and structural investigations. One of these aspects deals with the correlation among packing of molecules, the density and the stability of a certain crystal structure. For example, it is known that L-serine crystal is more dense than DL-serine, but the L-form undergoes a structural phase transition at lower pressure [3, 4]. In other words, the expected role played by the packing of molecules in the crystal is not so important, being more important for the problem the hydrogen bond interactions of the molecules.

Among the proteic amino acids the simplest are the aliphatic ones, compressing glycine (non-chiral), L-alanine, L-isoleucine, L-valine and L-leucine. For the first three amino acids previous investigations do not shown any structural phase transition under temperature changes [5 – 8], although a study involving birrefrigence and light depolarization measurements have shown some symmetry breaking around 220 K for L-alanine [9]. It is also important to state that differently from the temperature change experiments, there is evidence that L-alanine undergoes a structural phase transition at about 2.2 GPa under high pressure conditions [10]. For the other two aliphatic amino acid crystals the picture is as follows. A temperature investigation on L-valine crystal using Raman spectroscopy technique showed that the material undergoes a phase transition between 100 and 120 K [11]. For L-leucine crystal a previous work points to the occurrence of a phase transition at 80 $^{o}$C through differential scanning calorimetry measurements [12]; the same work presented an unpolarized Raman investigation on L-leucine performed in the spectral range between 100 and 1700 cm$^{-1}$ for two temperatures, 300 and 360 K.

The objective of this work is two-fold: (i) To present the polarized Raman spectra of L-leucine crystal through the entire spectral range of the normal modes, 50 – 3200 cm$^{-1}$, and give a tentative assignment of them; (ii) To present the temperature evolution of the Raman spectra of L-leucine crystal, giving particular attention to the observation changes which were associated to a structural phase transition undergone by the material.

## Experimental:

Single crystals of L-leucine were grown from aqueous solution containing powder from Sigma by the slow evaporation method at controlled temperature. They were obtained as colorless tiny platelets, similarly with L-valine [11] and L-isoleucine [8] crystals. The backscattering light was analyzed using a Jobin Yvon Triplemate 64000 micro-Raman system equipped with an $N_2$-cooled CCD detector. The slits were set for a 2 $cm^{-1}$ spectral resolution. The excitation source for the Raman experiments was 514.5 nm radiation from an argon ion laser. In order to obtain high temperature a home-made hot finger was utilized. The experiments were accomplished with heating of the sample.

## Results and Discussion:

L-leucine (in the inset of Fig. 1 the molecular structure is shown) crystallizes in a monoclinic lattice belonging to the $P2_1$ space group, $\beta = 86.2^o$. The conformations of the two leucine molecules are similar but not identical to each other and to that in DL-leucine. According to Harding and Howieson [13], the carboxyl and amino groups are hydrogen bonded in a double layer very like that in other non-polar L-amino acids (for example, L-valine [11] and L-isoleucine [8]). In terms of the irreducible representations (IRep) of the $C_2$ factor group, the normal modes are decomposed as $\Gamma = 132\ A + 132\ B$, and because there are two acoustic modes at the B IRep and one at the A IRep, the optical modes are $\Gamma_{op} = 131\ A + 130\ B$.

Fig. 1 presents the Raman spectra of L-leucine crystals for two scattering geometries, z(yy)z and z(xx)z, in the spectral region 50 – 700 $cm^{-1}$. The axes were defined according to the following convention: the z-axis was that perpendicular to the plane of the platelet and the y-axis was defined as the axis coincident with the longest dimension of the crystal; the x-axis was defined perpendicular to the y- and z-axes. In general terms, the bands observed with wavenumber lower than 150 $cm^{-1}$ are associated to the lattice modes of the crystal (lat.) and they can give interesting insights about the stability of the structure under changes of thermodynamic parameter as temperature and pressure. The band at 175 $cm^{-1}$ is tentatively associated to a torsion of $CO_2^-$ unit, $\tau(CO_2^-)$, and bands at 185 and 205 $cm^{-1}$, for the z(yy)z geometry, can be associated to torsions of CH, $\tau$(CH) [8]. In the L-leucine spectrum in the z(xx)z scattering geometry it is observed two bands at 245 and 288 $cm^{-1}$ that can be associated with out-of-plane vibration of CH, $\gamma$ (CH), and $CH_3$ torsion, $\tau(CH_3)$, respectively [14]. The low intense band at 332 $cm^{-1}$ is tentatively assigned as a NCC deformation, $\delta$(NCC) [11], while the bands at 352, 406, 445 and 460 $cm^{-1}$ are assigned as skeletal structure deformations, $\delta$(skel.) [14]. Finally, in the spectra of L-leucine presented in Fig. 1 it is observed an intense band at 536 $cm^{-1}$ which is associated with the rocking of $CO_2^-$ unit, $r(CO_2^-)$ [8, 14].

Fig. 2 presents the Raman spectra of L-leucine in the spectral region 700 – 1270 $cm^{-1}$. The band at 777 $cm^{-1}$ is associated to a $CO_2^-$ deformation, $\delta(CO_2^-)$ [14]. The two bands at 838 and 849 $cm^{-1}$ present an inversion of intensity for the two scattering geometries; they are assigned as out-of-plane vibration of $CO_2^-$, $\gamma$ $(CO_2^-)$, and rocking of $CH_3$ , $r(CH_3)$ [8]. The 900 – 1100 $cm^{-1}$ spectral region is characterized by bands associated to several CC and CN stretching vibrations, $\nu$(CC) and $\nu$(CN) [8, 11]. The band observed at 1131 $cm^{-1}$ is assigned as rocking of $NH_3^+$ unit, $r(NH_3^+)$, as well as the doublet at 1177 and 1187 $cm^{-1}$ (in the z(xx)z scattering geometry) [11]. The band at 1240 $cm^{-1}$ is tentatively assigned as torsion of $CH_2$, $\tau$ $(CH_2)$ [8].

The Raman spectra of L-leucine crystal in the 1280 – 1700 $cm^{-1}$ spectral region is presented in Fig. 3. This is a rich region, where many bands are observed, in particular, for the z(yy)z scattering geometry. Most of the bands in the region 1300 – 1375 $cm^{-1}$ are assigned as deformations of CH unit, $\delta$(CH) [8, 12]. Bands at 1391 and 1411 $cm^{-1}$ are associated to symmetric bending of $CH_3$, $\delta_S(CH_3)$, and peaks at 1458 and 1475 $cm^{-1}$ are associated to asymmetric bending of $CH_3$, $\delta_a(CH_3)$. In the z(yy)z scattering geometry the bands observed at 1560, 1585 and 1626 $cm^{-1}$ are associated to stretching vibrations of $CO_2^-$, $\nu$ ($CO_2^-$), and, possibly, to bending vibration of $NH_3^+$, $\delta(NH_3^+)$ [8, 11, 12]. It is worthwhile to mention a particular remark related to the 1550 – 1650 $cm^{-1}$ spectral region: observing the peaks appearing in the z(yy)z scattering geometry it is very clear that the band at 1626 $cm^{-1}$ has an intensity greater than the band at 1585 $cm^{-1}$ which is greater than the peak at 1560 $cm^{-1}$. The observation of the same spectral region in L-valine [8] and L-isoleucine [11] reveals the same relationship of intensities for the three peaks. Additionally, the Raman spectrum of L-leucine presents no peak in this region for the z(xx)z scattering geometry; the same is true for L-valine [8] and L-isoleucine [11].

Fig. 4 presents the Raman spectra of L-leucine crystal for two scattering geometries in the 2800 – 3100 $cm^{-1}$ spectral region. Bands due to the stretching vibrations of methylene, methyne and ammonium groups are expected to be observed in this high wavenumber region. However, as it is well known, the bands associated to the stretching vibrations of the $NH_3$ group present low intensity in the Raman spectrum. In general, the other bands appear with wavenumbers very similar to those of L-isoleucine crystal; the tentative assignment of these bands is given in Table 1. A final observation is that no band was observed for wavenumber higher than 3100 $cm^{-1}$, indicating that the crystal has grown as an anidrous form; in fact, when OH stretching vibrations are present, a large band centered at ~ 3400 $cm^{-1}$ is observed.

**High-temperature Raman spectra**

The interest in the study of the vibrational properties of organic and semi-organic substances varying temperature has grown in the last years mainly because of the possibility to shed light on the question of hydrogen bonds [15] and to help on the understanding of the phenomenon of polymorphism, mainly related to pharmaceuticals, pigments, and optical materials, among others [16]. Amino acid crystals, in particular, due to the fact that they are structures kept together through hydrogen and van der Waals bonds are expected to be unstable under great temperature variations. This is true, for example, for L-valine crystals that under low-temperature conditions undergo a phase transitiona at about 100 K [11]. For L-isoleucine crystals, another aliphatic proteic amino acid, the same temperature changes verified in the L-valine experiment (20 – 300 K) do not affect the stability of the structure [8]. L-alanine crystal, the smallest chiral aliphatic amino acid, when submitted to the same temperature conditions, seems to maintain the orthorhombic room-temperature structure, although the existence of a strong dynamic Jahn-Teller effect originated from the $NH_3^+$ charge-lattice coupling would explain a previouly misterious lattice instability at ~ 250 K [9].

In this section we discuss the effect of high temperature (T > 297 K) on the Raman spectra of L-leucine crystal. Fig. 5 shows the temperature evolution of the Raman spectra for all spectral regions in the z(yy)z scattering geometry. In order to facilitate the discussion we divided the spectra into four regions (Figs. 5(a) – 5(d)). Fig. 5(a) presents the bands appearing in the high-wavenumber region. The Raman spectra in this region show little changes, being observed only slight variations in the intensity of the bands. In fact, increasing temperature the two main effects on the $CH_2$ and $CH_3$ stretching region are: (i) a band at 2990 $cm^{-1}$, initially well separated from the most intense bands of lower wavenumber decreases intensity and appears as a shoulder of the bands of 2960/ 2971 $cm^{-1}$; (ii) bands at 2927 and 2939 $cm^{-1}$, which are seen as a band with large linewidth in the room temperature spectrum, begins to be observed as distinct bands in the spectrum at 323 K. For higher temperatures the two bands are clearly observed as separated ones.

Fig. 5(b) presents the temperature evolution of the Raman spectra of L-leucine crystal for the z(yy)z scattering geometry in the spectral region 1280 – 1750 $cm^{-1}$. Here, several aspects can be cited. First, most of the bands decrease intensity as occurs for the band observed at 1300 $cm^{-1}$ in the spectrum of room temperature. However, for the doublet 1318 and 1324 $cm^{-1}$ an interesting effect is observed: by increasing temperature the low-intensity band at 1324 $cm^{-1}$ decreases intensity in such a way that in the spectra of 323 – 343 K it is observed only as a shoulder of the other band. In the spectrum taken at 353 K the band originally at 1324 $cm^{-1}$ is not visible and, at the same time, the neighboring band (originally at 1318 $cm^{-1}$) seems to increase intensity relatively to the other bands of the spectrum. A possible explanation for this not conventional fact is that starting from the 353 K, the doublet becomes degenerated.

The low-intensity bands observed in the room temperature spectrum at 1365 and at 1441 $cm^{-1}$ are seen only as shoulders of the intense neighboring bands (at 1343 and 1458 $cm^{-1}$, respectively) when the sample is heated up to 393 K. Another observation in the spectra of Fig. 5(b) refers to the bands between 1550 and 1650 $cm^{-1}$. The band observed originally at 1560 $cm^{-1}$ has its intensity vanishing for T =353 K. Also, the bands at 1585 and 1626 $cm^{-1}$ change intensities when temperature varies from 297 to 413 K, similarly to what occurs with low-wavenumber bands of L-valine crystal at low temperature [11] and with low-wavenumber bands of L-alanine crystal at high pressure conditions [10].

Fig. 5(c) shows the temperature evolution of the Raman spectra of L-leucine crystal for the z(yy)z scattering geometry in the spectral region 700 – 1250 $cm^{-1}$. One interesting aspect is worth mentioning. A band at ~ 810 $cm^{-1}$ begins to be observed when temperature reaches 353 K, i.e., at room temperature there is no band between 700 and 830 $cm^{-1}$ and at that temperature a band appears. Also interesting is the fact that at room temperature it is observed a doublet at 919 and 926 $cm^{-1}$; when the sample is heated the bands lose intensity but remain as two distinct bands up to 343 K. However, when the L-leucine crystal is submitted to a temperature of 353 K the doublet originates one only band.

Fig. 5(d) presents the Raman spectra of L-leucine crystals for the z(yy)z scattering geometry in the region 50 – 700 $cm^{-1}$ for several temperatures. In the region 300 – 700 $cm^{-1}$ it is observed that all bands decrease intensities and their linewidths increase, as one expects. Bands at 175, 185 and 205 $cm^{-1}$ (marked by three down arrows), associated to torsional modes, are well visible in the spectrum at room temperature but, when temperature is increased, one observes that the band originally at 185 $cm^{-1}$ decreases intensity in such a way that at 353 K only two bands are present (they are marked by two arrows). The two bands remain up to the highest temperature of the experiments.

The region of the external modes (wavenumbers up to ~ 150 $cm^{-1}$), in particular, deserves special attention. We have shown that at ~ 353 K a series of changes appears in the internal modes region of L-leucine crystal. This can be an indication that some change in the structure is taking place at that temperature, although other phenomena can explain changes in bands associated to internal mode vibrations. One of the most known examples is related to L-alanine crystal where the splitting of the $NH_3^+$ torsional mode is observed at ~ 220 K, but being associated to small distortions of the $NH_3^+$ group, not to a structural phase transition [6]. On the contrary, taurine crystal is an example where changes in bands associated to internal modes (e.g. torsional vibration of CSH moiety) are effectively related to a structural modification [17]. For L-leucine crystal one observes that at T = 297 K three different bands (up to 150 $cm^{-1}$) are present in the spectrum. When the sample is heated the evolution of the two bands of lowest energy is such that at the highest temperature they appear as a large band. The most important effect, however, is observed for the band at 110 $cm^{-1}$, also marked by an arrow: its intensity continuously decreases and at about 353 K it goes to zero. The vanishing of this band associated to a lattice vibration can in a straightforward way be interpreted as a structural phase transition undergone by L-leucine crystal at 353 K.

As pointed out in Ref. [18], the molecules of L-leucine in the unit cell are organized in layers parallel to the *bc* face of the crystal. These layers present two kinds of interactions: on one side they interact via hydrogen bonds and on the other side they interact through the methyl groups of the leucine side chains. Because we have observed changes in the low-wavenumber region, it is possible that the phase transition involves the rupture of one hydrogen bond among the three possibilities of H bonds on amino N atoms with different molecules. The understanding of the correct mechanism which must be achieved by other measurements (X-ray and neutron diffractions) will be important to shed light on the interactions of alternating polypeptides that have similar organization to those of L-leucine molecules in their crystalline structure.

## Conclusions:

Raman spectra of L-leucine crystal were investigated for the interval range 50 – 3100 $cm^{-1}$ (for wavenumbers higher than 3100 $cm^{-1}$ and up to 3600 $cm^{-1}$ no band was observed) and a tentative assignment of the modes was given. The temperature evolution of the Raman spectra showed a series of modifications in the internal mode region at about 353 K. In this same temperature changes in the external mode region furnished evidence for a structural phase transition undergone by L-leucine crystal. The mechanism of the phase transition and the space group of the new high temperature phase is an open question which would be answered by X-ray or neutron diffraction experiments. However, this physical observation puts L-leucine in the same class of L-valine crystal among the aliphatic amino acid crystals which present at least one type of structural change. In another class it is possible to find L-alanine [6], D-alanine [19], D-valine [20] and L-isoleucine [8], which are stable under a very large range of temperature (at least for low-temperature conditions).

## Acknowledgements

We thank Dr. J. Ramos Gonçalves for a critical reading of the manuscript. Financial support from FUNCAP and CNPq is gratefully acknowledged.

## References

[1] Monaco SB, Davis LE, Velsko SP, Wang FT, Eimerl D, Zalkin A. *J. Cryst. Growth* 1987; **85**: 252.

[2] Eimerl D, Velsko S, Davis L, Wang F, Loiacono G, Kennedy G. *IEEE J. Quantum Electron.* 1989; **25**: 179.

[3] Kolesnik EN, Goryainov SV, Boldyreva EV. *Dokl. Phys. Chem.* 2005; **404**: 169.

[4] Moggach SA, Allan DR, Morrison CA, Parsons S, Sawyer L. *Acta Crystallogr. B* 2005; **61**: 58.

[5] Murli C, Thomas S, Venkateswaran S, Sharma SM. *Physica B.*2005; **364**: 233.

[6] Barthes M, Bordallo HN, Dénoyer F, Lorenzo J-E, Zaccaro J, Robert A, Zontone F. *Eur. Phys. J. B* 2004; **37**: 375.

[7] Lehmann MS, Koetzle TF, Hamilton WC. *J. Am. Chem. Soc.* 1972; **94**: 2657.

[8] Almeida FM, Freire PTC, Lima RJC, Remédios CMR, Mendes J, Melo FEA, *J. Raman Spectrosc*. 2006; **37**: 1296.

[9] Barthes M, Vik AF, Spire A, Bordallo HN, Eckert J. *J. Phys. Chem. A* 2002; **106**: 5230.

[10] Teixeira AMR, Freire PTC, Moreno AJD, Sasaki JM, Ayala AP, Mendes J, Melo FEA. *Solid State Commun.* 2000; **116**: 405.

[11] Lima JA., Freire PTC, Lima RJC, Moreno AJD, Mendes J, Melo FEA. *J. Raman Spectrosc*. 2005; **36**: 1076.

[12] Bougeard D. *Ber. Bunsen-Ges. Phys. Chem. Chem. Phys*. 1983; **87**: 279.

[13] Harding MM, Howieson RM. *Acta Cryst.* 1976; **B32**: 633.

[14] Pawlukojc A, Leciejewicz J, Natkaniec I. *Spectroc. Acta A* 1996; **52**: 29.

[15] Jeffrey GA. An Introduction to Hydrogen Bonding, Oxford University Press, New York, 1997.

[16] Bernstein J, Polymorphism in Molecular Crystals, Oxford University Press, Oxford, 2002.

[17] Lima RJC, Freire PTC, Sasaki JM, Melo FEA, Mendes J, Moreira RL. *J. Raman Spectrosc*. 2001; **32**: 751.

[18] Coll M, Solans X, Font-Altaba M, Subirana JA. *Acta Cryst*. 1986; **C42**: 599.

[19] Wilson CC, Myles D, Ghosh M, Johnson LN, Wang W. *New J. Chem*. 2005; **29**: 1318.

[20] Wang WQ, Gong Y, Wang ZM, Yan CH. *J. Struct. Chem*. 2003; **22**: 539.

## Caption for the Figures

Figure 1: Raman spectra of L-leucine crystal in the 50 – 700 $cm^{-1}$ spectral region in the z(yy)z and z(xx)z scattering geometries. Inset: Molecular structure of L-leucine.

Figure 2: Raman spectra of L-leucine crystal in the 700 – 1270 $cm^{-1}$ spectral region in the z(yy)z and z(xx)z scattering geometries.

Figure 3: Raman spectra of L-leucine crystal in the 1280 – 1700 $cm^{-1}$ spectral region in the z(yy)z and z(xx)z scattering geometries.

Figure 4: Raman spectra of L-leucine crystal in the 2800 – 3100 $cm^{-1}$ spectral region in the z(yy)z and z(xx)z scattering geometries.

Figure 5: Raman spectra of L-leucine crystal in the 50 – 3200 $cm^{-1}$ spectral region in the z(yy)z scattering geometry for several temperatures.

## Tables:

Table 1. Experimental wavenumbers from the Raman spectra of L-leucine crystal at room temperature and a tentative assignment of the modes.

| Raman | | | Raman | | |
|---|---|---|---|---|---|
| z(xx)z | z(yy)z | Assignment[a] | z(xx)z | z(yy)z | Assignment[a] |
| ($cm^{-1}$) | ($cm^{-1}$) | | ($cm^{-1}$) | ($cm^{-1}$) | |
| 72 | 65 | lat. | 1149 | - | |
| 87 | 87 | lat. | 1177 | 1175 | r($NH_3^+$) |
| - | 110 | lat. | 1187 | 1188 | r($NH_3^+$) |
| 125 | - | lat. | 1240 | - | $\tau$($CH_2$) |
| 175 | 175 | $\tau$($CO_2^-$) | 1300 | 1300 | |
| - | 185 | $\tau$(CH) | 1318 | 1318 | |
| - | 205 | $\tau$(CH) | 1346 | 1343 | $\delta$(CH) |
| 245 | - | $\gamma$ (CH) | 1354 | 1351 | |
| 288 | - | $\tau$($CH_3$) | - | 1365 | |
| 332 | 332 | $\delta$(NCC) | - | 1375 | |
| 352 | - | $\delta$(skel.) | 1390 | 1391 | $\delta_s$($CH_3$) |
| 406 | 405 | $\delta$(skel.) | 1411 | 1411 | $\delta_s$($CH_3$) |
| - | 445 | $\delta$(skel.) | 1444 | 1441 | - |
| 460 | 460 | $\delta$(skel.) | 1458 | 1458 | $\delta_a$($CH_3$) |
| 536 | 536 | r($CO_2^-$) | 1475 | 1475 | $\delta_a$($CH_3$) |
| 671 | 671 | w($CO_2^-$) | - | 1516 | - |
| 777 | - | $\delta$($CO_2^-$) | - | 1560 | $\nu$($CO_2^-$) |
| 838 | 838 | $\gamma$ ($CO_2^-$) | - | 1585 | $\nu$($CO_2^-$) |
| 849 | 849 | r($CH_3$) | - | 1626 | $\nu$($CO_2^-$) |
| 919 | 919 | $\nu$(CC) | 2871 | 2871 | $\nu_s$ ($CH_3$) |
| 926 | 926 | $\nu$(CC) | 2899 | 2900 | $\nu_s$ ($CH_3$) |
| 947 | 947 | $\nu$(CC) | 2906 | 2906 | $\nu$ ($CH_2$) |
| 965 | 965 | $\nu$(CC) | 2928 | 2927 | $\nu$ ($CH_2$) |
| 1004 | - | $\nu$(CC) | 2941 | 2939 | $\nu$ (CH) |
| 1032 | 1033 | $\nu$(CN) | 2959 | 2960 | $\nu$ (CH) |
| 1083 | 1082 | $\nu$(CN) | 2971 | 2971 | $\nu_a$ ($CH_3$) |
| 1131 | 1131 | r($NH_3^+$) | 2990 | 2990 | $\nu_a$ ($CH_3$) |

[a] lat., lattice vibration; $\tau$, torsion; $\delta$, bending; r, rocking; w, wagging; $\gamma$, out-of-plane vibration; $\nu$, stretching; skel., skeletal vibration; s, symmetric; a, asymmetric.

Figure 1

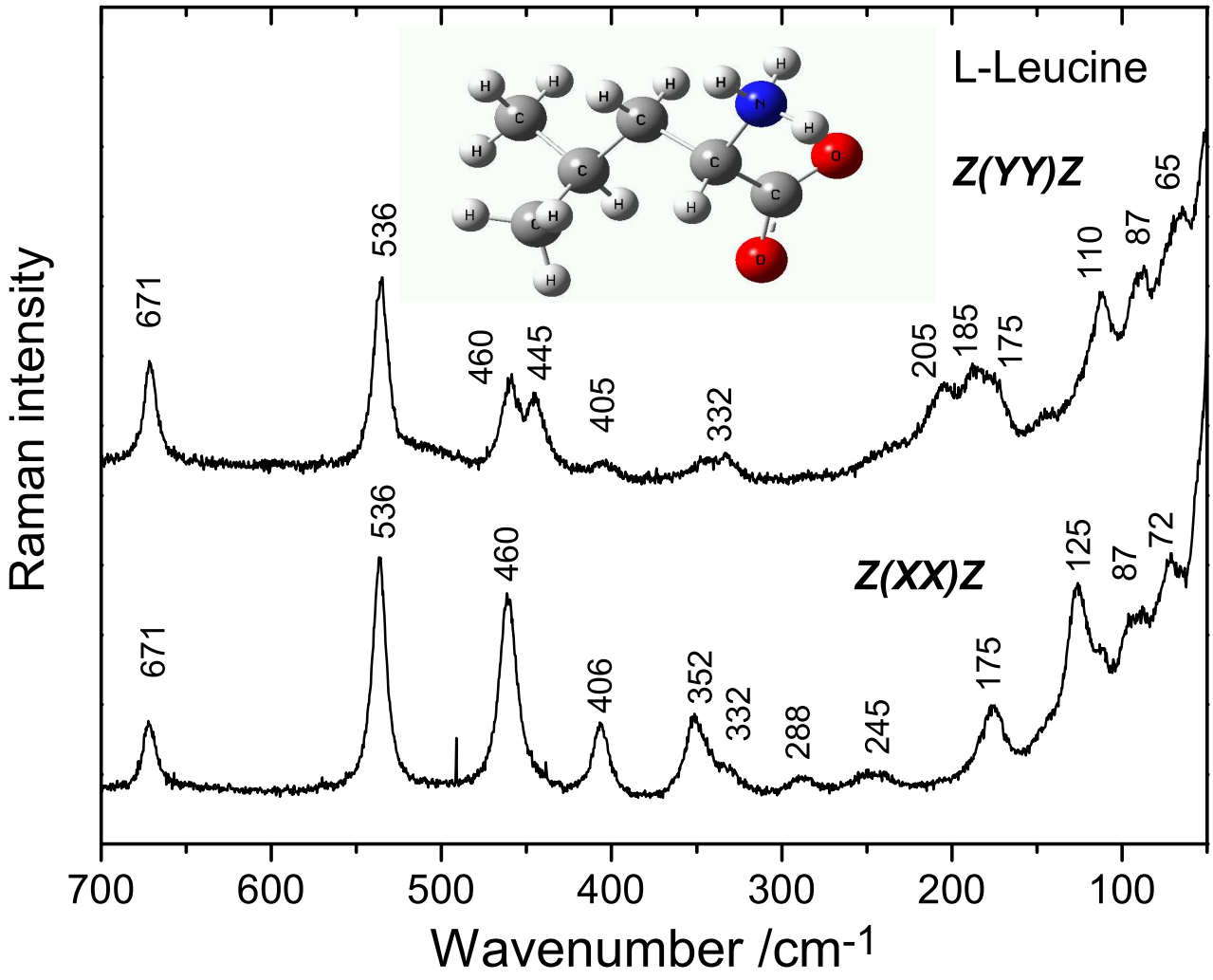

Figure 2

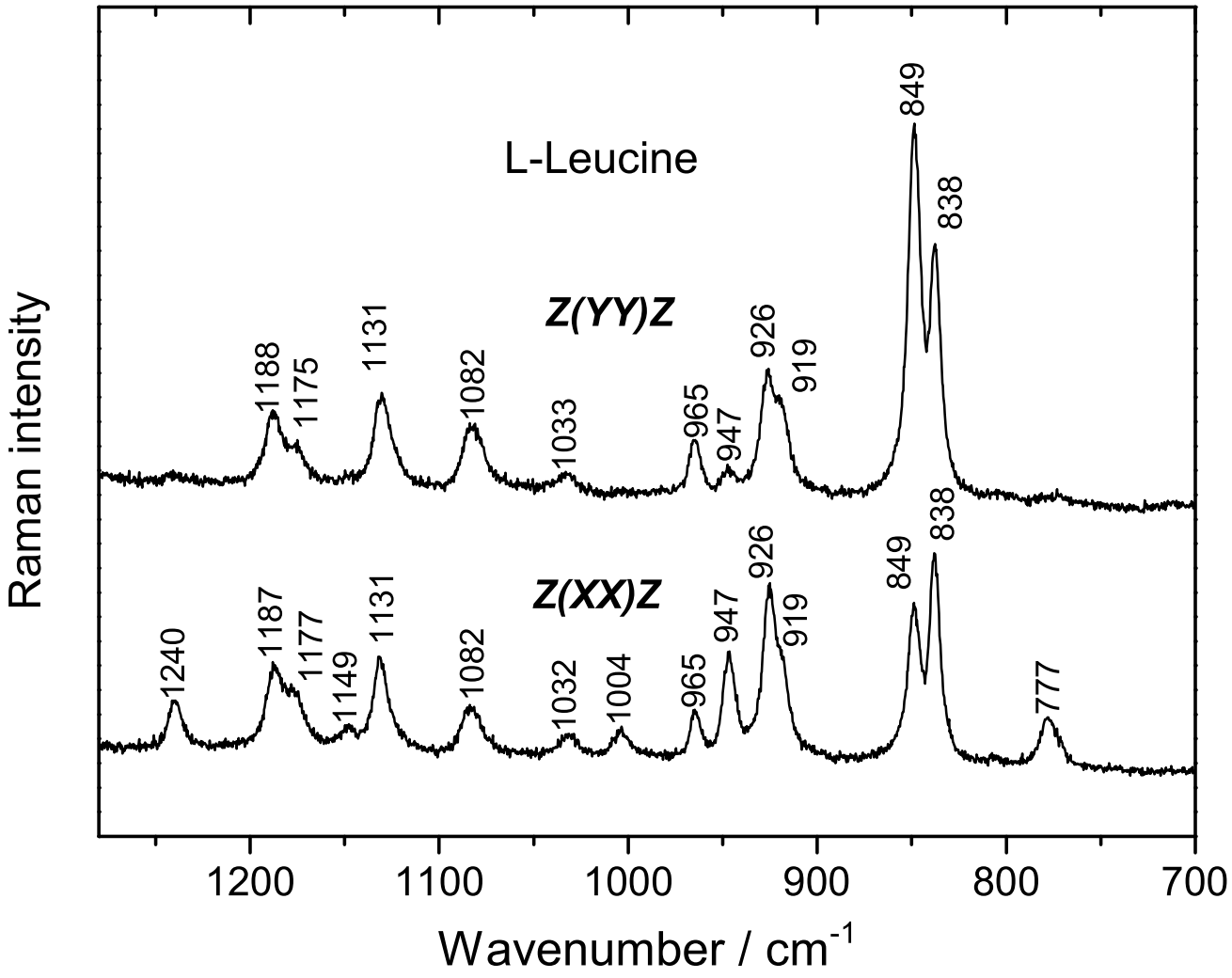

Figure 3

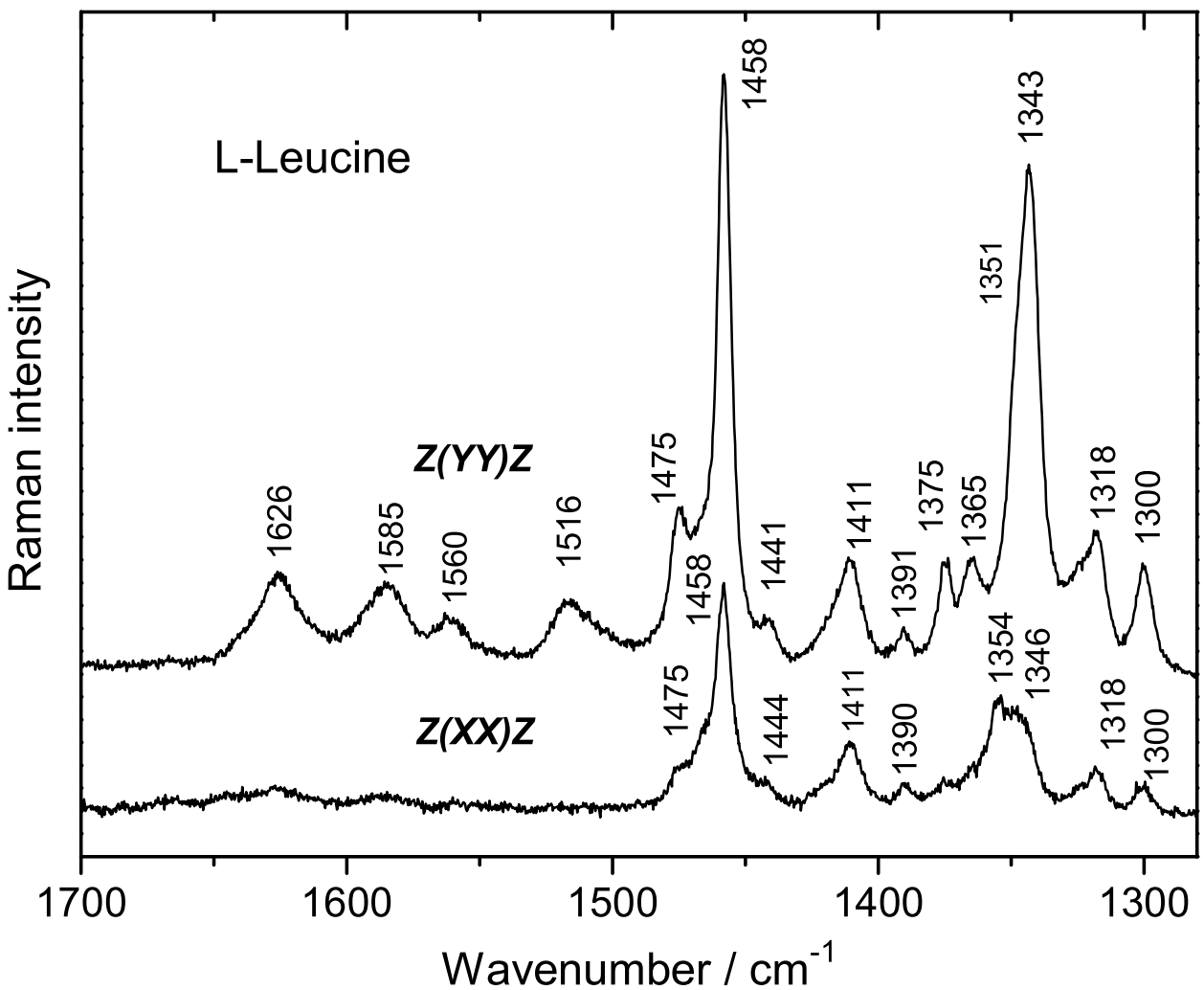

Figure 4

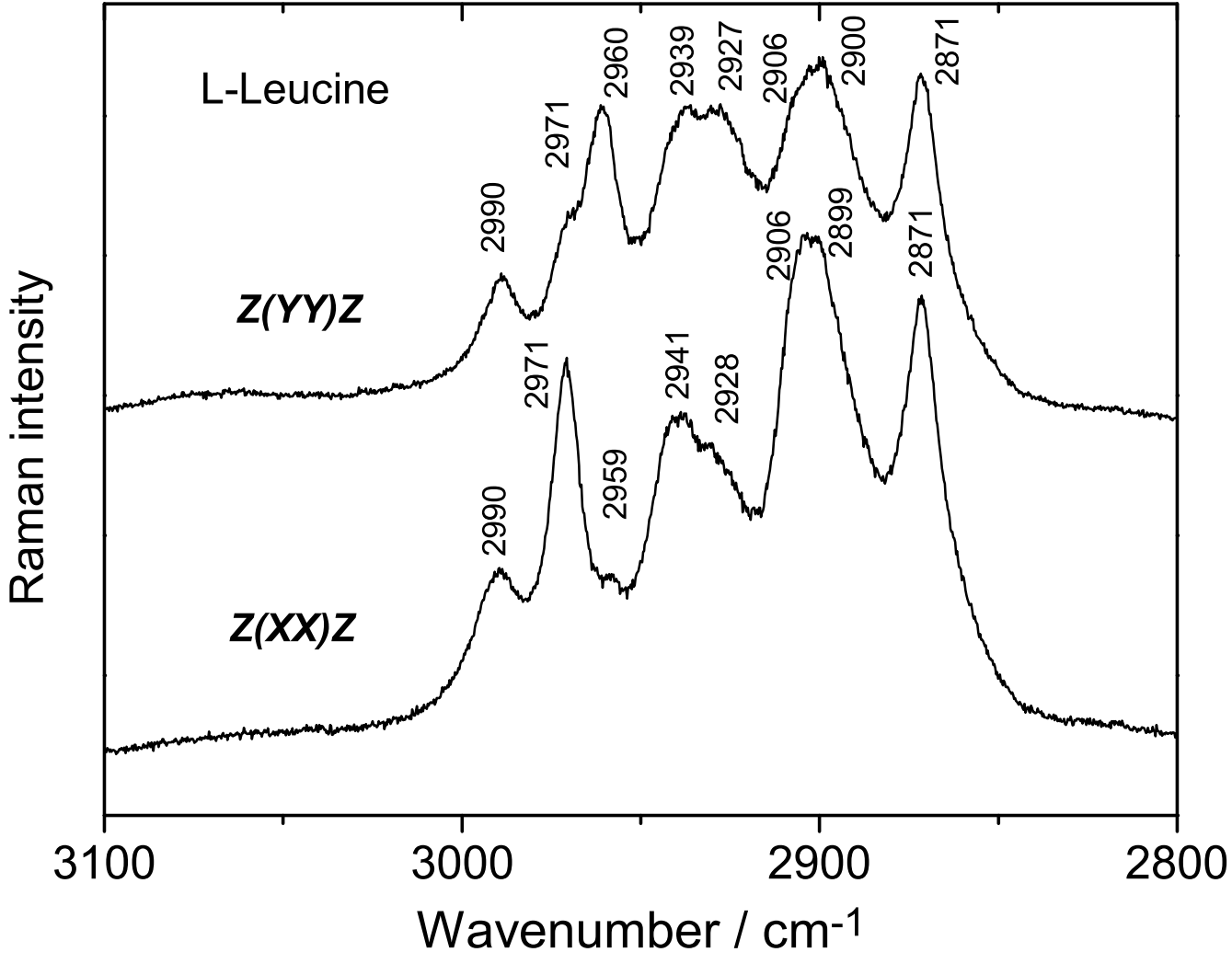

# Figure 5

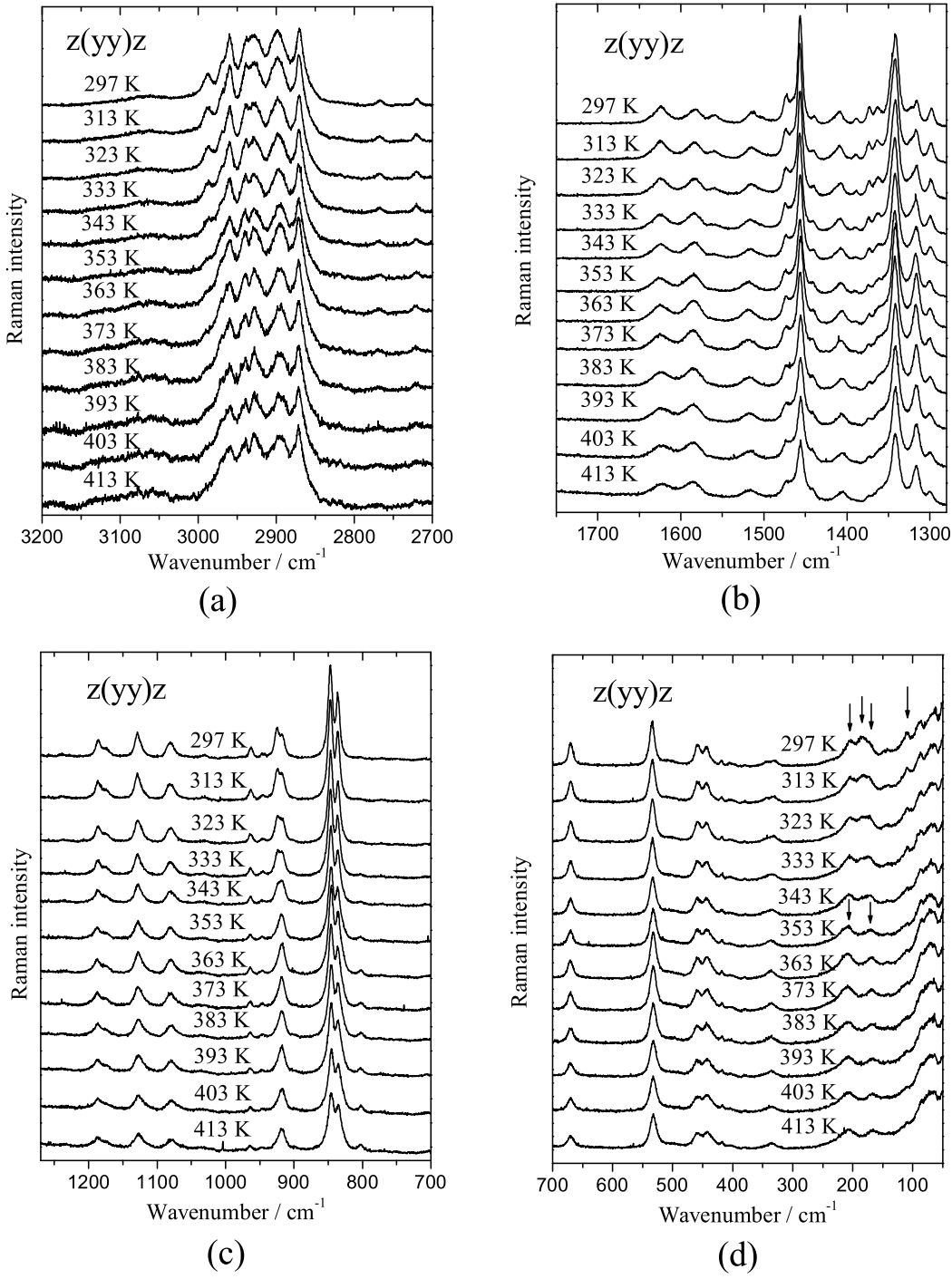